# Volatility Modeling of Stocks from Selected Sectors of the Indian Economy Using GARCH

Jaydip Sen
Department of Data Science
Praxis Business School
Kolkata, INDIA
email: jaydip.sen@acm.org

Sidra Mehtab
Department of Data Science
Praxis Business School
Kolkata, INDIA
email: smehtab@acm.org

Abhishek Dutta
Department of Data Science
Praxis Business School
Kolkata, INDIA
email: duttaabhishek0601@gmail.com

*Abstract*— Volatility clustering is an important characteristic that has a significant effect on the behavior of stock markets. However, designing robust models for accurate prediction of future volatilities of stock prices is a very challenging research problem. We present several volatility models based on generalized autoregressive conditional heteroscedasticity (GARCH) framework for modeling the volatility of ten stocks listed in the national stock exchange (NSE) of India. The stocks are selected from the auto sector and the banking sector of the Indian economy, and they have a significant impact on the sectoral index of their respective sectors in the NSE. The historical stock price records from Jan 1, 2010, to Apr 30, 2021, are scraped from the Yahoo Finance website using the DataReader API of the Pandas module in the Python programming language. The GARCH modules are built and fine-tuned on the training data and then tested on the out-of-sample data to evaluate the performance of the models. The analysis of the results shows that asymmetric GARCH models yield more accurate forecasts on the future volatility of stocks.

*Keywords—Volatility Clustering, GARCH, Residuals, Asymmetric GARCH, GJR-GARCH, EGARCH, Fixed Window Forecasting, Expanding Window Forecasting, MAE, RMSE.*

I. Introduction

While building robust and high precision models for accurate prediction of future stock has been considered a hot area of research, measuring and forecasting volatility in future stock values poses a more challenge. Interestingly, even though volatility in the stock market can present a significant risk to the investors, when correctly modeled and harnessed, it can be a source of substantial return in investments. Even if stock markets may fluctuate, crash, or surge, there is always a window of opportunity for profit if volatility in the market can be exploited. Classically, volatility is a measure of the dispersion of a return series of a stock, and it is measured by the standard deviation of the return series. The standard deviation of a return series of the stock shows how closely the return values from the stock are grouped or clustered around its mean value. If the return values are tightly coupled with each other over a period of time, the standard deviation, and hence the volatility of the series, is small. On the other hand, a high value of standard deviation implies significant volatility in the series. In a recent research work carried out by Crestmont Research, the relationship between the performance of stock markets and their volatility has been studied [1]. Based on the analysis of Standard & Poor's 500 Index (S&P500), Crestmont Research found that higher volatility in the stock market is usually associated with a higher probability of a declining index, while lower volatility is more likely to lead to a rise in the index of the stock market. This association of stock market volatility with its index value can be gainfully exploited by the investors in making long-term investment strategies for building their portfolios.

While there are numerous factors that can affect the volatility of stock markets, regional and economic factors such as taxation policies, monetary and fiscal policies, including interest rates, have a significant impact on the directional changes in the market, influencing the volatility by a substantial degree. In India, for example, when the central bank, the Reserve Bank of India, announces any change in the short-term interest rates for the banks, the volatility in the stock market immediately goes up. Higher volatility that is associated with bear markets usually leads to investors rebalancing their portfolios of stocks.

However, accurate forecasting of future volatility in the stock market is a very difficult task. Despite the availability of a large number of propositions and models in the literature, most of these models are found to work sub-optimally in real world applications leading to an inaccurate forecast of volatility. This paper presents a gamut of volatility models based on the approach of *generalized autoregressive conditional heteroscedasticity* (GARCH) [2]. We build the models using the historical stock price data in the NSE from Jan 1, 2010, to Apr 30, 2021. Ten stocks are chosen from two important sectors of the Indian economy- the auto sector and the banking sector. From each of these two sectors, the top five stocks are chosen, and several GARCH models are built, fine-tuned, and then backtested on the out-of-sample data. Extensive analysis is done on the evaluation of the performance of the volatility models.

The work has two major contributions. First, it proposes a set of GARCH models, which are built, fine-tuned, and backtested on real world stock price data over a period of more than ten years from different sectors of the Indian stock market. Second, it provides a benchmark of comparison of the volatility of the two sectors, which are studied in this work. Proper knowledge of the volatilities of the two sectors will enable a potential investor to understand the associated risk and return of the sectors.

We organize the paper as follows. Section II briefly presents some of the important recent works on volatility modeling. Section III present a discussion on how the relevant data are extracted and a step-by-step approach followed in the research methodology. Section IV presents extensive results on the volatility of two important sectors of the NSE of India. Finally, Section V concludes the paper.

II. Related Work

Researchers have done extensive works on the volatility analysis of stock markets and the design of robust portfolio systems based on the forecast of volatility and future stock prices using sophisticated predictive models [3-6]. A significant number of models based on the GARCH (1,1) framework have been proposed, and it has been observed that the generalized distribution of the residuals of these models is more accurate in assessing the volatility of the series than other models of residuals [7]. While some studies found that the asymmetric GJR-GARCH model produces more accurate forecasts on conditional variances when the volatility is higher, in most of the real-world scenarios, the EGARCH model yields more accurate forecasts in situations of asymmetric volatility [8]. It has been observed that GARCH-based volatility models yield more stable and robust forecasts while the sensitivity of entropy-based forecasts is higher [9]. Hence, the entropy-based models react faster in response to the availability of new information. The return analysis of fifteen important sectors of the Indian economy has been done based on the forecasted output of a deep learning-based *long-and-short-term memory* (LSTM) network model [10]. The study reveals that while the FMCG sector remains the most profitable one, the aggregate returns of the power sector are the lowest. Multivariate GARCH models have also been proposed for analyzing the volatilities of several contemporaneous time series [11].

## III. DATA AND METHODOLOGY

The methodology we followed in this work consists of ten steps. In the following, the steps are described in detail.

*(1) Data extraction:* We use the DataReader API from the *panadas* module to extract the stock price data from the Yahoo Finance website. The parameters passed to the DataReader function are the *ticker name* of the stock, the *website* from which the records are extracted, the *start* date, the *end* date, and the *attributes* of the data extracted. For example, for extracting Maruti Suzuki stock price records listed in the NSE from the Yahoo Finance website from the date *start* to the date *end*, with the attributes *high*, *low*, *open*, *close*, *volume,* and *adjusted close*, the required python code is: *maruti = web.DataReader ('MARUTI.NS', 'yahoo', start, end)[['High', 'Low', 'Open', 'Close', 'Adj Close']]*. For all the stocks, we use the start date as Jan 1, 2010, and the end date as Apr 30, 2021.

*(2) Hurst exponent and volatility computation:* Once the stock price records are extracted, we compute the *Hurst exponent* of the time series of the *close* prices. The Hurst exponent is a measure of the long-term behavior of a time series. It measures the autocorrelations among the values at different lags in a time series and computes the rates at which the autocorrelations decrease with the lag values. Based on these computations, Hurst exponent calculates a measure that determines whether a time series regresses strongly to a mean value or clusters towards a direction, upward to downward. A value of the Hurst exponent between 0 to 0.5 indicates a time series exhibiting long-term switching between successive high and low values. For these types of time series, a high value is usually followed by a low value. A value of 0.5 for the Hurst exponent implies a time series consisting of uncorrelated values. For these types of time series, the autocorrelations decay exponentially and quickly go down to zero. Time series with Hurst exponent values between 0.5 to 1 have long-term positive correlations, wherein a high value is likely to be followed by another high value, and the long-term future values also tend to be higher. We compute the Hurst exponent for all stocks using a function that computes the autocorrelations at all possible lags in the series with a maximum lag of 100.

After computing the Hurst exponent of a stock based on its *close* index, we compute the volatility of the series. The daily volatility is determined by computing the standard deviation of the daily return values. The monthly and the annual volatility values are computed by multiplying the daily volatility by a factor of the *square root* of 21 and the *square root* of 252, respectively, assuming that there are 21 working days in a month and 252 working days in a year. For computing the standard deviation of a series, the *std* function defined in Python is used.

*(3) Study of the statistical properties of the return series and the log return series:* In this step, we compute the return series and the log return series. For obtaining the return series, we use the *pct_change* function in Python. The log return series is computed by first applying the *log* function on the *close* values of the stock. The log function is defined in the *NumPy* library. The successive differences of the log values over different days are computed using the *diff* function. Finally, the results of the *diff* function are multiplied by 100 to obtain the percentage of *log return* values. We plot the return series, the log return series, their Q-Q (Quantile-Quantile) plots, the *autocorrelation function* (ACF) plots, and the *partial autocorrelation function* (PACF) plots. For plotting these graphs, we use a max lag value of 100 so that significant lags for the ACF and PACF plots can be identified, even if the lag values are high. The Q-Q plots reveal the important statistical measures of the series like mean, standard deviation, skewness, and kurtosis.

*(4) Fitting a GARCH (1,1) model with a constant mean and normally distributed residuals:* In this step, we build volatility models for the stock price time series. First, we build a GARCH (1,1) with a constant mean and normally distributed residuals. A GARCH model of this type can be expressed as (1) in which $\varepsilon_{t-1}$ and $\sigma_{t-1}$ represent the residual and the variance at the time instant *t*-1, respectively, $\sigma_t$ represents that variance at time instant *t*. The first term, ω represents a constant variance that corresponds to the long-run average volatility, the coefficient α represents the effect of the square of the residual value at time instant *t*-1, and the coefficient β represents the effect of the variance at time instant *t*-1 on the volatility at time instant *t*, i.e., $\sigma_t^2$.

$$\sigma_t^2 = \omega + \alpha\varepsilon_{t-1}^2 + \beta\sigma_{t-1}^2 \qquad (1)$$

The GARCH model uses residuals as the volatility shocks. For the GARCH (1,1) model at this step, we assume that the residuals follow a normal distribution with a zero mean. The residual at time instant *t*, $\varepsilon_t$ is given by (2). In (2) $r_t$ represents the return at time instant *t* and $\mu_t$ is the mean value of the return series.

$$\varepsilon_t = r_t - \mu_t \qquad (2)$$

Using the *arch_model* function defined in the *arch* module of Python, we fit a GARCH (1,1) model with constant mean and normally distributed residuals. Using the *summary* function of the model, we determine the values of the parameters ω, α, and β of the GARCH model and their respective *p*-values. The *Akaike information criteria* (AIC) of the model is also found out. The standardized residuals and the conditional volatility of the return series are plotted as predicted by the GARCH model. The standardized residuals are the residual values normalized by the volatility values.

*(5) Fitting a GARCH (1,1) model with a constant mean and skewed-t distributed residuals:* Since the stock return and the log return values usually do not follow a normal distribution, we fine-tune the GARCH model built in Step 4 by assuming the residuals to follow a skewed *t* distribution rather than a normal distribution. A *t*-distribution is symmetric around its mean like a normal distribution, but since it has a much wider spread of its two tails, it can accommodate values that are far away from the mean value. We also plot the volatility of the GARCH (1,1) model with skewed-*t* error distribution with respect to the daily return of the stock to verify how accurately the GARCH model is able to capture the pattern of variations of the daily return series.

*(6) Identifying the best-fit ARMA model for the log return series:* In this step, we find the best-fit *autoregressive moving average* (ARMA) model for the log return series of the stock. We use the *auto_arima* function defined in the *pmdarima* module of Python for finding the best-fit model. The model that yields the lowest value for *Bayesian information criteria* (BIC) is the optimum one. We also note the number of iterations needed to find the best-fit model.

*(7) Fitting the ARMA residuals into a GARCH (1,1) model:* The best ARMA model is fitted on the log return series, and the residuals of the model are computed using the *resid* attribute of the model. A GARCH (1,1) model with zero mean (as the residuals are assumed to have a zero mean) is then fitted into the residuals of the ARMA model. Using the *summary* function of the GARCH model, we find out the parameters ω, α, and β and their corresponding *p*-values. The AIC of the model is also noted. The significance levels of the *p*-values show the *goodness of fit* of the GARCH model into the ARMA model's residuals.

*(8) Fitting asymmetric volatility models on the return series and evaluating the goodness of fits of the models:* The GARCH models built in the previous steps assume that positive and negative news have a similar impact on the volatility. This assumption does not hold good in the real world, where the impact of negative news on the volatility of return is more than the impact of positive news. To effectively model the volatility of the return series, GARCH models must have the capability of handling the asymmetric nature of the impact. We build two asymmetric models for this purpose, (i) GJR-GARCH and (ii) EGARCH.

The GJR-GARCH (1,1,1) model is given by (3). In (3), γ is an asymmetric parameter and $d_{t-1}$ is a dummy variable. When $\varepsilon_{t-1} < 0$, $d_{t-1} = 1$, and when $\varepsilon_{t-1} > 0$, $d_{t-1} = 0$. In other words, if the residual in the previous time instant was negative, the dummy variable $d_{t-1}$ is assigned a value of 1, while a positive residual value in the previous time instant implies a zero value for the dummy variable.

$$\sigma_t^2 = \omega + \alpha\varepsilon_{t-1}^2 + \beta\sigma_{t-1}^2 + \gamma\varepsilon_{t-1}^2 d_{t-1} \quad (3)$$

The EGARCH (1,1,1) model is given by (4)

$$\log(\sigma_t^2) = \omega + \beta g(Z_{t-1}) + \alpha \log(\sigma_{t-1}^2) \quad (4)$$

In (4), the value of the function $g(Z_t)$ is evaluated using $g(Z_t) = \theta Z_t + \lambda(|Z_t| - E(|Z_t|))$ and $Z_t = \frac{\varepsilon_t}{\sigma_t}$.

If $\theta = 0$, then larger shocks increase the conditional variance if $(|Z_t| - E(|Z_t|) > 0$, else the conditional variance is decreased. If $\theta < 1$, the shock in variance $g(Z_t)$ is positive if the shocks $Z_t$ are less than the mean. Therefore, the negative shocks in returns cause the shock to the conditional variance to be positive if $\theta$ is much less than 1.

Both GJR-GARCH (1,1,1) and EGARCH (1,1,1) models are implemented using the *arch_model* function in Python with the parameters *p*=1, *q*=1, and *o*=1, and the error distribution set to *t*. The *summary* functions of the models are invoked to determine the values of the parameters ω, α, γ, and β, and their respective *p*-values. We plot the return series with the forecasted volatility values of GJR-GARCH and EGARCH models to check how close the forecasted volatility values are with respect to the actual volatility values of the return series.

*(9) Testing the accuracy of the EGARCH volatility forecasting:* In this step, we use two different methods of forecasting, the *expanding window method* and the *fixed window method*, to compute the forecasted volatility using the EGARCH model on the *out-of-sample* data from Jan 1, 2021, to Apr 30, 2021. In the fixed window method, based on the previous five days' volatility values, the volatility values for the next five days are forecasted. In this case, the size of the training window (i.e., five days) remains constant. However, for the expanding window method, the size of the training window increases with the number of rounds of forecasting. Finally, the forecasted volatility values versus the actual volatility of the return series are plotted.

*(10) Backtesting the EGARCH models on the out-of-sample data:* In the final step, we carry out backtesting of the EGARCH model on the return series and compute the *mean absolute error* (MAE), and the *root mean square error* (RMSE) of the model.

## IV. EXPERIMENTAL RESULTS

This section presents extensive experimental results on the performance of the various volatility models based on GARCH.

For the purpose of our study, we select two important sectors of stocks listed in the NSE. These two sectors are the auto sector and the banking sector. We identify the top five stocks from each of these two sectors based on the NSE *indexogram* reports [12]. We use the Python programming language version 3.7.4 for implementing the GARCH models in this work. Using the *DataReader* function of the *pandas_datareader* module, the historical stock price records are extracted from Jan 1, 2010, to Apr 30, 2021, from the Yahoo Finance website. The GARCH models are built using the *arch_model* function defined in the *arch* module of Python. The orders of the ARMA models are determined using the *auto_arima* function defined in the *pmdarima* module in Python. The models are trained and tested on a system with an Intel i7-9750H processor, having a clock speed of 2.60GHz-2.59 GHz, 16 GB RAM, running on the operating system of Windows 10. We present extensive results for the two sectors in the following.

*Auto sector:* The top five stocks of the auto sector listed in the NSE are: (i) Maruti Suzuki India (MSZ), (ii) Mahindra and Mahindra (MAH), (iii) Tata Motors (TAM), (iv) Bajaj Auto (BAJ), and (v) Hero MotoCorp (HMC). The respective percentages of weights of these five stocks used in computing the auto sector index are as follows: MSU-18.72, MMH-15.72, TMO-11.50, BAJ-10.89, HMC-7.99 [12].

The Hurst exponents and the daily, monthly, and annual volatility values of the stocks are presented in Table I. Due to space constraints, we present the visuals for only the topmost stock from each sector. Figure 1 presents some important characteristics of the log return series of the Maruti Suzuki stock. Figure 2 depicts the properties of the daily volatility series of the Maruti Suzuki stock. From Table 1, it is evident that the stocks BAJ, MAH, and HMC exhibit mean-reverting characteristics, while the series of MSZ and TAM show a mild mean-reverting property based on their Hurst exponent values. The annual volatility (which is used as the measure of the volatility of a series) is the highest for TAM, while BAJ exhibits the lowest value of annual volatility.

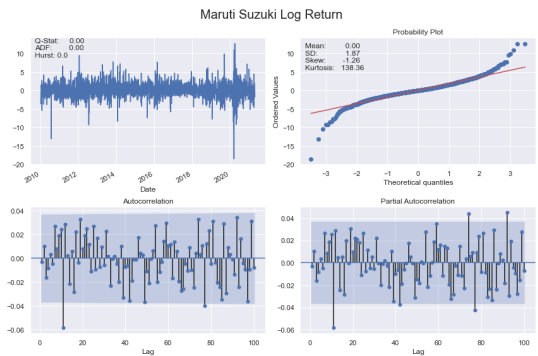

Fig. 1. Maruti Suzuki daily log return: (a) the daily log return plot, (b) the Q-Q plot, (c) the ACF plot, (d) the PACF plot.

TABLE I. HURST EXP. AND VOLATILITY OF AUTO SECTOR STOCKS

| Stock | Hurst Exp | Daily Vol | Moth. Vol | Ann. Vol |
|---|---|---|---|---|
| BAJ | 0.4508 | 1.66 | 7.62 | 26.41 |
| MSZ | 0.5104 | 1.87 | 8.55 | 29.62 |
| TAM | 0.5221 | 2.73 | 12.51 | 43.32 |
| MAH | 0.4854 | 1.96 | 8.96 | 31.05 |
| HMC | 0.4410 | 1.87 | 8.56 | 29.64 |

Tables II and III present important statistical properties of the log return series and the daily volatility of the log return series of the auto sector stocks. From Table II, it is evident that while the mean values of the log return series of all the stocks are zero, the standard deviation, the skewness, and the kurtosis for TAM are the highest. The same observation is noted in Table III. It is, therefore, concluded that TAM is the most volatile stock in the auto sector. BAJ, on the other hand, is found to exhibit the lowest volatility in this sector. Both ACF and PACF plots for the daily log return series and the daily volatility series are found to be significant over large values of lags. We plotted the ACF and the PACF graphs for a max value of lag of 100 and found that these plots are significant over large lags. These are evident from Figures 1 and 2.

Next, we fit a GARCH (1,1) model on the return series using a constant mean, volatility model as GARCH, and the error distribution as normal. We use the *arch_model* function with parameters, *p*=1, *q*=1, *mean=constant*, *vol=GARCH*, and *dist=normal*. Table IV presents the results for the GARCH (1,1) model fitting on the return series of the five stocks of the auto sector. The columns in Table IV show respectively the name of the stocks (*Stock*), the number of iterations needed to fit the model into the return series data (*#Iterations*), the number of times the function (i.e., the *log-likelihood function*) is evaluated (*Func Eval*), the number of times the gradients are updated (*Grad Eval*), and the value of the log-likelihood function (*Log-Likelihood*).

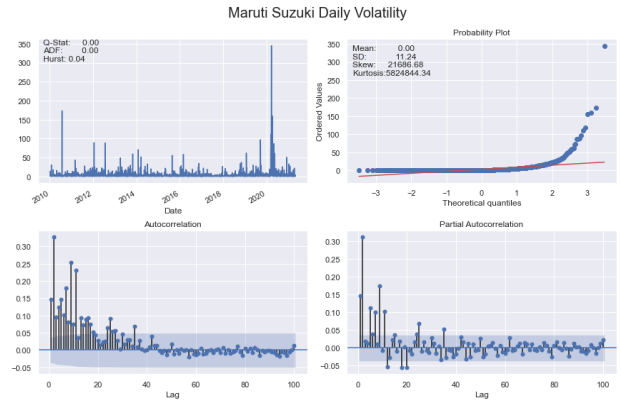

Fig. 2. Maruti Suzuki daily volatility: (a) the daily volatility plot, (b) the Q-Q plot, (c) the ACF plot, (d) the PACF plot

TABLE II. LOG RETURN SERIES PROPERTIES OF AUTO SECTOR STOCKS

| Stock | Mean | SD | Skew | Kurt | ACF (q) | PACF (p) |
|---|---|---|---|---|---|---|
| BAJ | 0.00 | 1.66 | -0.49 | 67.49 | 85 | 85 |
| MSZ | 0.00 | 1.87 | -1.26 | 138.36 | 80 | 90 |
| TAM | 0.00 | 2.70 | 9.06 | 671.83 | 100 | 100 |
| MAH | 0.00 | 1.95 | 1.32 | 104.24 | 100 | 100 |
| HMC | 0.00 | 1.86 | 2.03 | 102.87 | 100 | 100 |

TABLE III. LOG RETURN DAILY VOLATILITY SERIES OF AUTO SECTOR

| Stock | Mean | SD | Skew | Kurt | ACF (q) | PACF (p) |
|---|---|---|---|---|---|---|
| BAJ | 0.00 | 7.74 | 6108 | 997913 | 40 | 80 |
| MSZ | 0.00 | 11.24 | 21686 | 5824844 | 40 | 40 |
| TAM | 0.00 | 24.87 | 341563 | 286164821 | 100 | 100 |
| MAH | 0.00 | 9.48 | 9193 | 1642584 | 90 | 90 |
| HMC | 0.00 | 9.53 | 10860 | 2176305 | 50 | 60 |

TABLE IV. GARCH (1,1) MODEL FITTING FOR AUTO SECTOR STOCKS

| Stock | # Iterations | Func Eval | Grad Eval | Log Likelihood |
|---|---|---|---|---|
| BAJ | 10 | 64 | 10 | -5272.09 |
| MSZ | 13 | 79 | 13 | -5515.69 |
| TAM | 16 | 98 | 16 | -6603.94 |
| MAH | 12 | 73 | 12 | -5652.86 |
| HMC | 11 | 67 | 11 | -5534.73 |

TABLE V. AUTO SECTOR GARCH (1,1) MODEL WITH NORMAL ERROR

| Stock | mu | | omega | | alpha[1] | | beta[1] | | AIC |
|---|---|---|---|---|---|---|---|---|---|
| | coeff | p-val | coeff | p-val | coeff | p-val | coeff | p-val | |
| BAJ | 0.078 | 0.008 | 0.101 | 0.054 | 0.049 | 0.005 | 0.914 | 0.000 | 10552.2 |
| MSZ | 0.103 | 0.001 | 0.159 | 0.164 | 0.058 | 0.018 | 0.892 | 0.000 | 11039.4 |
| TAM | 0.045 | 0.322 | 0.036 | 0.543 | 0.024 | 0.299 | 0.972 | 0.000 | 13215.9 |
| MAH | 0.073 | 0.025 | 0.052 | 0.029 | 0.046 | 0.000 | 0.941 | 0.000 | 11313.7 |
| HMC | 0.036 | 0.257 | 0.153 | 0.136 | 0.071 | 0.016 | 0.885 | 0.000 | 11077.5 |

Table V presents the summary of the GARCH (1,1) model with constant mean, and normally distributed error. The parameter *mu* refers to the constant mean, *omega* refers to the constant variance that corresponds to the *long-run average volatility*, *alpha* is the new information (i.e., the innovation) in the current round that was not available in the previous round, and *beta* refers to the forecasted volatility of the previous period. In simple words, *alpha* is the coefficient of the *square of the residuals* in the previous round, and beta is the coefficient associated with the *variance of the previous round*. The higher the value of *alpha*, the bigger is the impact of the shock. On the other hand, the larger the value of *beta*, the longer is the shock duration. It is observed in Table V, the value of *alpha* is the highest for MSZ, while for TAM, it is the lowest. Hence, the impact of a shock (i.e., a sudden change) is the most in MSZ, while that is the least for the TAM stock. On the other hand, based on the values of *beta*, the duration of shock is found to be the longest for TAM, while it is the shortest for HMC.

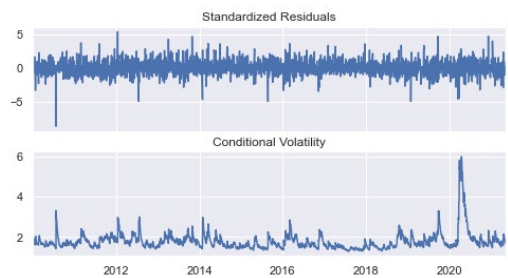

Fig. 3. Maruti Suzuki standardized residuals and conditional volatility of the constant mean GARCH model with normal error

Figure 3 exhibit the patterns of the standardized residuals and the conditional volatility of the MSZ stock as computed by the GARCH (1,1) model built on the assumption of a constant mean and normally distributed residuals.

TABLE VI. BEST FIT ARMA MODELS FOR AUTO SECTOR STOCKS

| Stock | Best ARMA Model | AIC | BIC |
|---|---|---|---|
| BAJ | (0, 0, 0) (0, 0, 0) No Intercept | 10751.78 | 10757.70 |
| MSZ | (0, 0, 0) (0, 0, 0) No Intercept | 11402.67 | 11408.60 |
| TAM | (0, 0, 0) (0, 0, 0) No Intercept | 12470.07 | 13476.03 |
| MAH | (0, 0, 0) (0, 0, 0) No Intercept | 11647.40 | 11653.33 |
| HMC | (0, 0, 0) (0, 0, 0) No Intercept | 11381.74 | 11387.68 |

TABLE VII. GARCH MODEL ON ARMA RESIDUALS FOR AUTO SECTOR

| Stock | omega | | alpha[1] | | beta[1] | | # Obs | AIC |
|---|---|---|---|---|---|---|---|---|
| | coeff | p-val | coeff | p-val | coeff | p-val | | |
| BAJ | 0.101 | 0.051 | 0.050 | 0.006 | 0.913 | 0.000 | 2788 | 10545.40 |
| MSZ | 0.172 | 0.197 | 0.057 | 0.032 | 0.889 | 0.000 | 2790 | 11044.70 |
| TAM | 0.029 | 0.379 | 0.019 | 0.115 | 0.977 | 0.000 | 2790 | 13195.00 |
| MAH | 0.054 | 0.024 | 0.046 | 0.000 | 0.940 | 0.000 | 2790 | 11302.20 |
| HMC | 0.149 | 0.139 | 0.070 | 0.014 | 0.886 | 0.000 | 2790 | 11057.10 |

Next, we change the error model from normal to *skewed-t*. In the financial time series, the residuals are not normal most of the time, and hence, the assumption of the residuals following a *skewed-t* distribution is more realistic. We use the *arch_model* function to fit a GARCH (1,1) model with the parameter *dist=skewt* and keeping all other parameters the same as in the GARCH (1,1) model built earlier with the assumption of normally distributed residuals. Figure 4 depicts the GARCH model *volatility* with *skewed-t distributed residuals* vs. the *daily return* values for the MSZ stock. It is evident that the GARCH model is able to capture the volatility pattern of the return series very accurately.

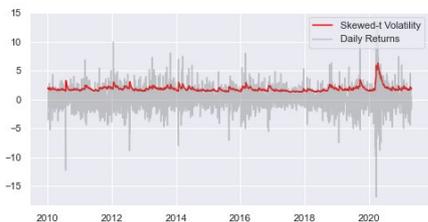

Fig. 4. Maruti Suzuki volatility of GARCH model with *skewed-t* distribution error vs. daily return

To further improve the GARCH (1,1) model, next, we model the log-returns of the *close* prices of the stock with an ARMA model and then fit the residuals of the ARMA models to estimate the volatility of the log return series with a new GARCH model. For this purpose, we use the *auto_arima* function defined in the *pmdarima* module of Python. Using the *auto_arima* function and BIC as the information criteria, we identify the best-fit ARMA model for the log return series of the stocks. Table VI presents the results. It is observed that for all five stocks, ARMA (0,0,0) (0,0,0) with no intercept is the best-fit model. The residuals of this ARMA are then fitted into a *zero-mean* GARCH (1,1) model. Table VII presents the results for the five stocks. It is observed that the *p*-values of most of the *alpha* and the *beta* coefficients are significant, indicating that the residuals have fitted very well into the constant mean GARCH model.

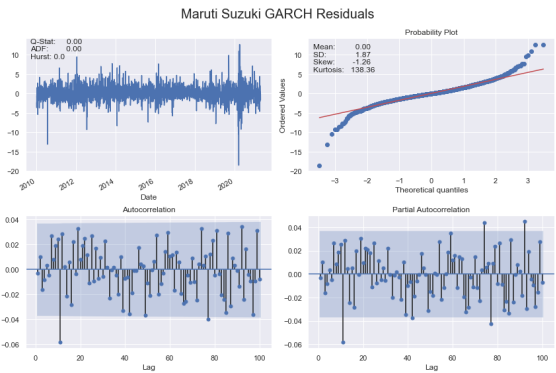

Fig. 5. Maruti Suzuki GARCH residual with the fitted ARMA model

TABLE VIII. CONSTANT MEAN GJR-GARCH MODEL ON AUTO SECTOR

| Stock | mu | | omega | | alpha | | gamma | | beta | |
|---|---|---|---|---|---|---|---|---|---|---|
| | coef | pval | coef | pval | coef | pval | coef | pval | coef | pval |
| BAJ | 0.017 | 0.51 | 0.11 | 0.09 | 0.01 | 0.24 | 0.08 | 0.01 | 0.92 | 0.00 |
| MSZ | 0.025 | 0.37 | 0.07 | 0.01 | 0.01 | 0.21 | 0.08 | 0.00 | 0.93 | 0.00 |
| TAM | 0.010 | 0.82 | 0.08 | 0.37 | 0.01 | 0.35 | 0.04 | 0.07 | 0.96 | 0.00 |
| MAH | 0.029 | 0.34 | 0.05 | 0.01 | 0.02 | 0.05 | 0.05 | 0.00 | 0.94 | 0.00 |
| HMC | -0.020 | 0.49 | 0.08 | 0.26 | 0.03 | 0.24 | 0.04 | 0.01 | 0.93 | 0.00 |

TABLE IX. CONSTANT MEAN EGARCH MODEL ON AUTO SECTOR

| Stock | mu | | omega | | alpha | | gamma | | beta | |
|---|---|---|---|---|---|---|---|---|---|---|
| | coef | pval | coef | pval | coef | pval | coef | pval | coef | pval |
| BAJ | 0.012 | 0.71 | 0.03 | 0.02 | 0.10 | 0.00 | -0.06 | 0.00 | 0.97 | 0.00 |
| MSZ | 0.022 | 0.44 | 0.03 | 0.00 | 0.09 | 0.00 | -0.06 | 0.00 | 0.98 | 0.00 |
| TAM | -0.021 | 0.63 | 0.30 | 0.20 | 0.09 | 0.01 | -0.04 | 0.00 | 0.99 | 0.00 |
| MAH | 0.021 | 0.50 | 0.03 | 0.00 | 0.11 | 0.00 | -0.05 | 0.00 | 0.98 | 0.00 |
| HMC | -0.024 | 0.40 | 0.04 | 0.06 | 0.15 | 0.00 | -0.02 | 0.07 | 0.97 | 0.00 |

For further improving the accuracy of the GARCH models, we build two asymmetric models using the concepts of GJR-GARCH and EGARCH. The asymmetric models more effectively capture the volatility of financial times series of the real world. We first fit a GJR-GARCH model on the return series of the auto sector stocks. The *arch_model* function is called with the following parameters: *p*=1, *q*=1, *o*=1, *vol*=GARCH, and *dist*=t. This creates a GJR-GARCH (1,1,1) model having a constant mean value, residuals following a *t* distribution, and the volatility with an asymmetric GARCH pattern. The asymmetric shock considered in this model is of one lag. Table VIII shows the model summary of the five stocks of the auto sector. It is found that almost all the *gamma* and the *beta* coefficients are significant. Thus, the one-lag asymmetric shocks and one-lag variances are found to be the most significant predictors in the GJR-GARCH volatility model of the auto sector stocks.

Finally, we build another asymmetric volatility model using EGARCH (1,1,1). Unlike other GARCH models, EGARCH does not have any requirement of the *alpha* and the *beta* parameters to be non-negative. Hence, EGARCH model building requires less amount of time. We use the *arch_model* function with the parameter *vol*=EGARCH and the rest of the values of the parameters identical to what has been used in the GJR-GARCH model. Table IX presents the summary of the EGARCH models for the auto sector stocks. The model is found to be a very good fit on the return values of the stocks as almost all *alpha*, *gamma*, and *beta* coefficients are significant. Table X presents the BIC values for the GJR-GARCH and EGARCH models for the auto sector stocks. Except for MAH, the BIC for the EGARCH model is found to be lower than that of the corresponding GJR-GARCH model. Figure 6 exhibits the return series of MSZ, the GJR-GARCH volatility, and the EGARCH volatility. It is evident, the performance of both the GARCH models are almost identical, and both the models have accurately captured the volatility of the return series of MSZ.

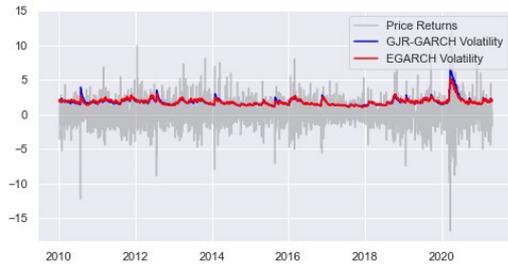

Fig. 6. Maruti Suzuki plot of daily return vs. GJR-GARCH-fitted volatility and EGARCH-fitted volatility

TABLE X. GJR-GARCH AND EGARCH MODEL BIC FOR AUTO SECTOR

| Stock | GJR-GARCH BIC | EGARCH BIC |
|---|---|---|
| BAJ | 10335.27 | 10334.19 |
| MSZ | 10754.15 | 10739.47 |
| TAM | 12940.56 | 12920.59 |
| MAH | 11186.22 | 11188.57 |
| HMC | 10880.33 | 10873.16 |

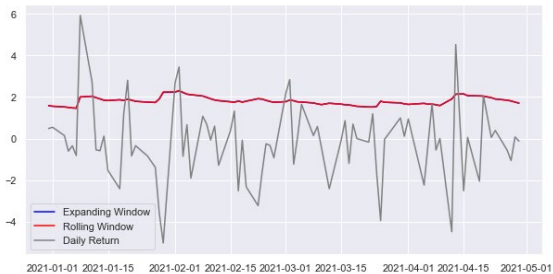

Fig. 7. Maruti Suzuki GARCH forecasts – fixed window forecast, expanding window forecast vs. daily return (Period: Jan 1, 2021 –May 1, 2021)

Using the *expanding rolling window method* and the *fixed rolling window method*, we forecast the volatility of the MSZ return series from Jan 1, 2021, for 80 data points. Figure 7 exhibits the results. It is evident that both the forecasting methods have yielded identical forecasts. It is also clearly seen that the models have been able to forecast the volatility of the return series quite effectively.

TABLE XI. PERFORMANCE OF THE EGARCH MODEL ON AUTO SECTOR

| Stock | MAE | RMSE |
|---|---|---|
| BAJ | 3.06 | 7.37 |
| MSZ | 3.85 | 10.27 |
| TAM | 8.14 | 29.50 |
| MAH | 4.12 | 9.51 |
| HMC | 3.86 | 9.81 |

To evaluate the performance of the EGARCH model on the return series of the auto sector stocks, we carry out backtesting of the model. Table XI presents the mean *absolute error* (MAE) and the *root mean square error* (RMSE) of the model for all five stocks. For computing the MAE and the RMSE values, we use the *mean_absolute_error* and the *mean_squared_error* functions defined in the *metric* submodule of the *sklearn* module. The reason for the high MAE and RMSE values for TAM is due to the high volatility of the stock.

*Banking sector:* The five stocks that contribute most significantly in the computation of the banking sector index of the NSE and their respective weights in percentage figures are: (i) HDFC Bank (HDB) – 27.41, (ii) ICICI Bank (ICB) – 21.09, Axis Bank (AXB) – 14.30, Kotak Mahindra Bank (KMB) – 13.02, State Bank of India (SBI) – 11.74 [12].

Table XII presents the Hurst exponent, the daily, monthly, and annual volatility of the five stocks of the banking sector. It is evident that AXB, ICB, and SBI exhibit a very mild trending nature, while HDB and KMB are mean-reverting types. While AXB is the most volatile stock, HDB exhibits the least volatility among the five stocks.

TABLE XII. HURST EXP. AND VOLATILITY OF BANKING SECTOR STOCKS

| Stock | Hurst Exp | Daily Vol | Moth. Vol | Ann. Vol |
|---|---|---|---|---|
| AXB | 0.5161 | 2.32 | 10.61 | 36.77 |
| HDB | 0.4926 | 1.50 | 6.89 | 23.88 |
| ICB | 0.5048 | 2.18 | 10.00 | 34.65 |
| KMB | 0.4895 | 1.83 | 8.41 | 29.12 |
| SBI | 0.5093 | 2.22 | 10.15 | 35.16 |

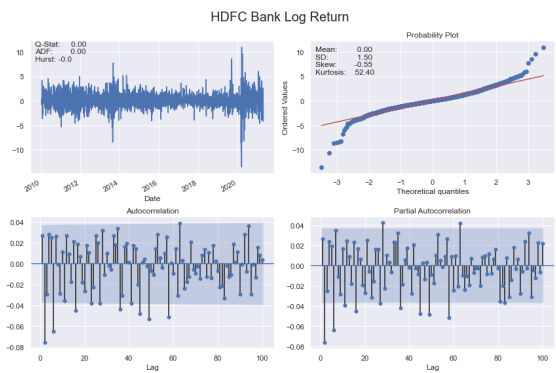

Fig. 8. HDFC Bank daily log return: (a) the daily log return plot, (b) the Q-Q plot, (c) the ACF plot, (d) the PACF plot.

Table XIII and Figure 8 present some important statistical properties of the log return series of the banking sector stocks. While all return series have zero mean values, the SD of the AXB stock is found to be the highest. The magnitude of the skewness and the kurtosis of AXB are also found to be the highest among all the five stocks of the banking sector indicating that it is the most volatile one among the banking sector stocks. The significant lags for the ACF and the PACF plots of the log return series of all the stocks are found to be very high. For plotting the ACF and PACF, a maximum lag value of 100 has been used.

Table XIV and Figure 9 depict the characteristics of the volatility of the log return series of the stocks. Again, we find that the mean values of all the volatility values of the daily log return are zero. AXB exhibits the highest value for the SD, the skewness, and the kurtosis of the volatility of the log return series. The volatility of SBI also shows high values of skewness and kurtosis. The significant lags for the ACF and the PACF plots are found to be high for the volatility series as well.

TABLE XIII. LOG RETURN SERIES PROPERTIES OF BANKING SECTOR

| Stock | Mean | SD | Skew | Kurt | ACF (q) | PACF (p) |
|---|---|---|---|---|---|---|
| AXB | 0.00 | 2.33 | -9.55 | 609.54 | 100 | 95 |
| HDB | 0.00 | 1.50 | -0.55 | 52.40 | 65 | 65 |
| ICB | 0.00 | 2.18 | 0.15 | 184.19 | 80 | 80 |
| KMB | 0.00 | 1.83 | -0.39 | 84.21 | 40 | 40 |
| SBI | 0.00 | 2.20 | 5.18 | 265.97 | 95 | 95 |

TABLE XIV. LOG RETURN DAILY VOLATILITY OF BANKING SECTOR

| Stock | Mean | SD | Skew | Kurt | ACF (q) | PACF (p) |
|---|---|---|---|---|---|---|
| AXB | 0.00 | 24.09 | 460932 | 472592483 | 40 | 40 |
| HDB | 0.00 | 6.88 | 4338 | 584839 | 40 | 45 |
| ICB | 0.00 | 12.72 | 28823 | 8842304 | 60 | 60 |
| KMB | 0.00 | 8.54 | 6328 | 885018 | 60 | 60 |
| SBI | 0.00 | 15.58 | 85234 | 45649397 | 20 | 20 |

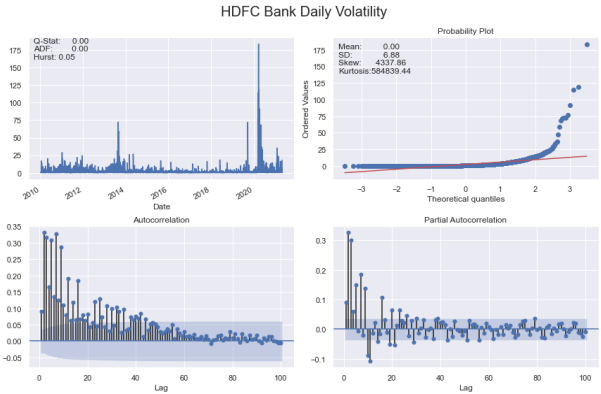

Fig. 9. HDFC Bank daily volatility: (a) the daily volatility plot, (b) the Q-Q plot, (c) the ACF plot, (d) the PACF plot

TABLE XV. GARCH (1,1) MODEL FITTING FOR BANKING SECTOR

| Stock | # Iterations | Func Eval | Grad Eval | Log Likelihood |
|---|---|---|---|---|
| AXB | 10 | 62 | 10 | -6031.98 |
| HDB | 12 | 72 | 12 | -4776.36 |
| ICB | 13 | 77 | 13 | -5983.41 |
| KMB | 12 | 71 | 12 | -5445.46 |
| SBI | 9 | 56 | 9 | -6057.60 |

As in the case of the auto sector stocks, we fit a constant-mean GARCH (1,1) model on the return series with normally distributed residuals. The arch_model function is used with the parameters, *p*=1, *q*=1, mean=*constant*, vol=*GARCH*, and dist=*normal*. Table XV presents results of the model fitting on the five stocks of the banking sector. The significance of the column names of Table XV is the same as that of Table IV, which has already been explained earlier.

TABLE XVI. BANKING SECTOR GARCH (1,1) WITH NORMAL ERROR

| Stock | mu | | omega | | alpha[1] | | beta[1] | | AIC |
|---|---|---|---|---|---|---|---|---|---|
| | coeff | p-val | coeff | p-val | coeff | p-val | coeff | p-val | |
| AXB | 0.118 | 0.002 | 0.173 | 0.002 | 0.077 | 0.000 | 0.889 | 0.000 | 12072.0 |
| HDB | 0.113 | 0.000 | 0.034 | 0.021 | 0.062 | 0.000 | 0.923 | 0.000 | 9560.7 |
| ICB | 0.111 | 0.003 | 0.243 | 0.001 | 0.080 | 0.000 | 0.869 | 0.000 | 11974.8 |
| KMB | 0.114 | 0.000 | 0.109 | 0.056 | 0.074 | 0.000 | 0.892 | 0.000 | 10898.9 |
| SBI | 0.021 | 0.625 | 0.582 | 0.047 | 0.159 | 0.061 | 0.736 | 0.000 | 12123.2 |

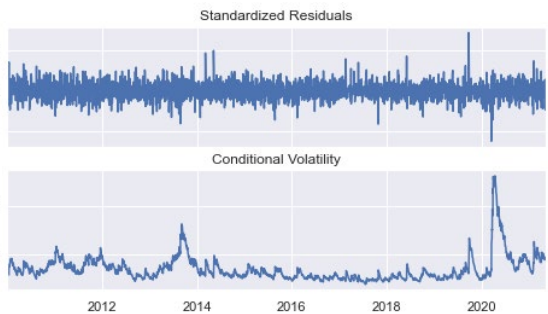

Fig. 10. HDFC Bank standardized residuals and conditional volatility of the constant mean GARCH model with normal error

The summary of the *constant-mean* GARCH (1,1) model with normally distributed residuals for the banking sector stocks is presented in Table XVI. As explained earlier, the parameters *mu*, *omega*, *alpha*, and *beta* refer to the constant mean, the long-term constant variance, the coefficient of the square of the last round residual value, and the coefficient of the variance of the last round of volatility forecasting. A high value of *alpha* causes more impact of a shock on the volatility, while a large *beta* value causes the impact to remain effective over a longer period of time. SBI is found to have the highest value of *alpha*, while the parameter value is the lowest for HDB. Hence, the impact of a shock is the highest for SBI, and it is the least for HDB. On the basis of the *beta* values, HDB has the longest duration of the impact of a shock, while SBI has the least duration of an impact.

The standardized residuals and the conditional volatility of the HDB stock return series as computed by the GARCH (1,1) model are shown in Figure 10. This GARCH model is built based on two assumptions: (i) a constant mean of the return series and (ii) normally distributed residuals.

TABLE XVII. BEST FIT ARMA MODELS FOR BANKING SECTOR STOCKS

| Stock | Best ARMA Model | AIC | BIC |
|---|---|---|---|
| AXB | (0, 0, 0) (0, 0, 0) No Intercept | 10751.78 | 12642.05 |
| HDB | (0, 0, 0) (0, 0, 0) No Intercept | 10204.09 | 10210.03 |
| ICB | (0, 0, 0) (0, 0, 0) No Intercept | 12265.54 | 12271.47 |
| KMB | (0, 0, 0) (0, 0, 0) No Intercept | 11304.02 | 11309.95 |
| SBI | (0, 0, 0) (0, 0, 0) No Intercept | 12316.88 | 12322.82 |

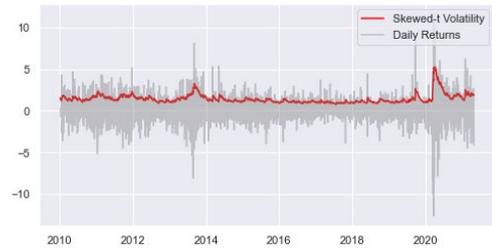

Fig. 11. HDFC Bank volatility of GARCH model with *skewed-t* distribution error vs. daily return

As we did for the auto sector stocks, to make our models more realistic, we build a GARCH model with the assumption of the residuals following a *skewed-t* distribution instead of a normal distribution. In building this model, the function *arch_model* is used with the parameter *dist=skewt*. The other parameters are kept identical to those used in the GARCH model with normally distributed residuals. Figure 11 depicts the GARCH model *volatility* with *skewed-t distributed residuals* and the daily return values of the HDB stock. It is seen that the volatility model very closely followed the daily return volatility of the stock.

TABLE XVIII. BANKING SECTOR GARCH MODEL ON ARMA RESIDUALS

| Stock | omega | | alpha[1] | | beta[1] | | # Obs | AIC |
|---|---|---|---|---|---|---|---|---|
| | coeff | p-val | coeff | p-val | coeff | p-val | | |
| AXB | 0.185 | 0.002 | 0.080 | 0.000 | 0.883 | 0.000 | 2790 | 12068.50 |
| HDB | 0.034 | 0.020 | 0.061 | 0.000 | 0.923 | 0.000 | 2790 | 9565.48 |
| ICB | 0.230 | 0.000 | 0.077 | 0.000 | 0.874 | 0.000 | 2790 | 11947.50 |
| KMB | 0.113 | 0.058 | 0.075 | 0.000 | 0.890 | 0.000 | 2790 | 10891.70 |
| SBI | 0.512 | 0.030 | 0.137 | 0.021 | 0.765 | 0.000 | 2790 | 12068.70 |

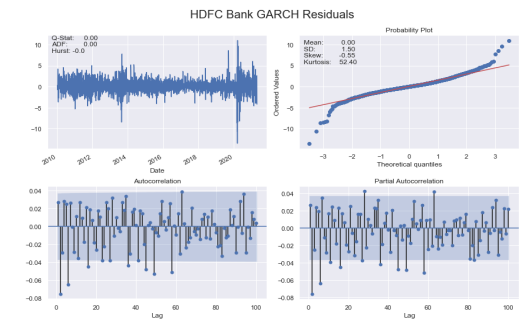

Fig. 12. HDFC Bank GARCH residual with the fitted ARMA model

Next, we fine-tune the GARCH (1,1) model. We build an ARMA model using the log-returns of the *close* prices of the stocks. The residuals of the ARMA model are fitted into a GARCH model to estimate the long-term volatility of the log return series. The *auto_arima* function defined in the *pmdarima* module of Python is used to identify the order of the ARMA model. With the BIC as the information criteria, the order of the best-fit ARMA model is determined. This ARMA model is fitted on the log return series of the stocks. Table XVII depicts the results. Similar to the auto sector stocks, for all the five stocks of the banking sector, the ARMA (0,0,0) (0,0,0) with no intercept is found to be the best-fit model. A zero-mean GARCH (1,1) model is fitted into the residuals of the ARMA model. The summary of the GARCH model is presented in Table XVIII. Most of the coefficients of *omega*, *alpha*, and *beta* are found to have significant *p*-values indicating a very good model fit.

TABLE XIX. CONSTANT MEAN GJR-GARCH ON BANKING SECTOR

| Stock | mu | | omega | | alpha | | gamma | | beta | |
|---|---|---|---|---|---|---|---|---|---|---|
| | coef | pval | coef | pval | coef | pval | coef | pval | coef | pval |
| AXB | 0.033 | 0.34 | 0.12 | 0.00 | 0.02 | 0.07 | 0.09 | 0.00 | 0.92 | 0.00 |
| HDB | 0.057 | 0.01 | 0.02 | 0.02 | 0.02 | 0.08 | 0.07 | 0.00 | 0.94 | 0.00 |
| ICB | -1.182 | 0.00 | 0.09 | 0.00 | 0.00 | 1.00 | 0.12 | 0.00 | 0.93 | 0.00 |
| KMB | 0.07 | 0.01 | 0.06 | 0.02 | 0.03 | 0.03 | 0.06 | 0.00 | 0.93 | 0.00 |
| SBI | 0.011 | 0.75 | 0.29 | 0.00 | 0.06 | 0.01 | 0.05 | 0.01 | 0.86 | 0.00 |

TABLE XX. CONSTANT MEAN EGARCH ON BANKING SECTOR

| Stock | mu | | Omega | | alpha | | gamma | | beta | |
|---|---|---|---|---|---|---|---|---|---|---|
| | coef | pval | coef | pval | coef | pval | coef | pval | coef | pval |
| AXB | 0.023 | 0.51 | 0.04 | 0.00 | 0.13 | 0.00 | -0.07 | 0.00 | 0.98 | 0.00 |
| HDB | 0.049 | 0.02 | 0.11 | 0.00 | 0.10 | 0.00 | -0.06 | 0.00 | 0.99 | 0.00 |
| ICB | -0.016 | 0.63 | 0.03 | 0.00 | 0.11 | 0.00 | -0.09 | 0.00 | 0.98 | 0.00 |
| KMB | 0.067 | 0.00 | 0.02 | 0.00 | 0.12 | 0.00 | -0.05 | 0.00 | 0.98 | 0.00 |
| SBI | 0.004 | 0.90 | 0.08 | 0.01 | 0.16 | 0.00 | -0.03 | 0.03 | 0.95 | 0.00 |

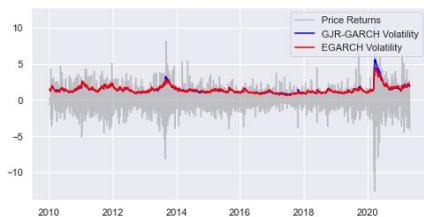

Fig. 13. HDFC Bank plot of daily return vs. GJR-GARCH-fitted volatility and EGARCH-fitted volatility

As in the case of auto sector stocks, for making the volatility model more accurate, we build two asymmetric models, EJR-GARCH and EGARCH. The asymmetric models more effectively capture the volatility of financial times series of the real world. First, the GJR-GARCH model is built using the *arch_model* function with the parameters *p*=1, *q*=1, *o*=1, *vol*=GARCH, and *dist*=t. A constant-mean GJR-GARCH model is created whose residuals follow a *t* distribution, with one-lag asymmetric shock as significant. The summary of the model is presented in Table XIX. Since all *gamma* and *beta* coefficients are significant, it is clear that the one-lag asymmetric shocks and one-lag variances are the dominating components of the model. The results of the EGARCH (1,1) model in Table XX indicates a better model fit that the GJR-GARCH (1,1) since all the coefficients are significant here. From Table XXI, it is evident that the BIC of the EGARCH model is smaller than that of GJR-GARCH for all stock excepts AXB, indicating its better fit into the stock price data. Figure 14 exhibits the return series of HDB and the volatilities of the GJR-GARCH and the EGARCH. Table XXII presents the mean *absolute error* (MAE) and the *root mean square error* (RMSE) of the EGARH model for all five stocks of the banking sector. It is evident that the model has performed very well on the out-of-sample data.

TABLE XXI. GJR-GARCH AND EGARCH BIC FOR BANKING SECTOR

| Stock | GJR-GARCH BIC | EGARCH BIC |
|---|---|---|
| AXB | 11901.02 | 11908.21 |
| HDB | 9433.43 | 9428.42 |
| ICB | 11764.83 | 11763.56 |
| KMB | 10764.76 | 10761.22 |
| SBI | 11870.66 | 11866.09 |

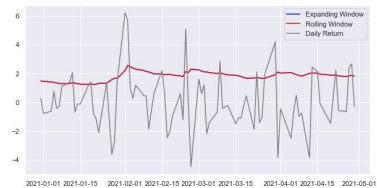

Fig. 14. HDFC Bank forecasts – fixed window forecast, expanding window forecast vs. daily return (Period: Jan 1 –May 1, 2021)

TABLE XXII. PERFORMANCE OF EGARCH MODEL ON BANKING SECTOR

| Stock | MAE | RMSE |
|---|---|---|
| AXB | 5.67 | 19.42 |
| HDB | 2.34 | 6.25 |
| ICB | 5.00 | 11.70 |
| KMB | 3.51 | 7.96 |
| SBI | 5.27 | 17.81 |

V. CONCLUSION

In this paper, we have proposed several volatility models based on different variants of GARCH. The models are built on the historical stock price data from Jan 1, 2010, to Apr 30, 2021. The stocks are chosen from the auto sector and the banking sector of the NSE of India. The models are fine-tuned and then backtested on the out-of-sample data to estimate their accuracy in the prediction of future volatility of the stocks. While it is observed that asymmetric GARCH models outperform their symmetric counterparts, EGARCH is found to have yielded the most accurate results.